\documentclass[12pt,a4paper]{article}
\usepackage{amsfonts}
\usepackage{amsthm}
\usepackage{amssymb}
\usepackage{epsfig}

\begin{document}

\begin{center}
{\bf\Large Modelling Quintessential Inflation in Palatini Modified Gravity}

\bigskip

{\large Konstantinos Dimopoulos,$^a$}
{\large Alexandros Karam,$^b$}

{\large Samuel S\'anchez L\'opez$^a$}
{\large and Eemeli Tomberg$^b$}%
\footnote{
\begin{tabular}{l}
k.dimopoulos1@lancaster.ac.uk,\\
alexandros.karam@kbfi.ee,\\
eemeli.tomberg@kbfi.ee,\\
s.sanchezlopez@lancaster.ac.uk
\end{tabular}
}

\bigskip

$^a${\em Consortium for Fundamental Physics, Physics Department,\\Lancaster University, Lancaster LA1 4YB, United Kingdom.}

$^b${\em Laboratory of High Energy and Computational Physics, 
National Institute of Chemical Physics and Biophysics, R{\"a}vala pst.~10, Tallinn, 10143, Estonia}

\end{center}

\begin{abstract}
  We study a model of quintessential inflation constructed in $R^2$ modified
  gravity with a non-minimally coupled scalar field, in the Palatini
  formalism. Our non-minimal inflaton field is characterised by a simple
  exponential potential. We find that successful quintessential inflation can
  be achieved with no fine-tuning on the model parameters. Predictions on
  the characteristics of dark energy will be tested by observations in the near
  future, while contrast with existing observations provides insights on the
  modified gravity background, such as the value of the non-minimal coupling
  and its running.
\end{abstract}

\section{Introduction}

Observations suggest that the Universe has undergone at least two phases of
accelerated expansion. The primordial phase is called cosmic inflation and it
is responsible for arranging the fine tuning needed for the subsequent
hot big bang evolution of the Universe, as well as for generating the
cosmological perturbations, which are responsible for structure formation
\cite{Lythbook}. The late phase is taking place at present and it is attributed
to the gradual dominance of the mysterious dark energy substance, which makes
up almost 70\% of the Universe content today \cite{DEbook}.

In the context of fundamental theory, cosmic inflation is typically realised
according to the inflationary paradigm, which suggests that the expansion of
the Universe is accelerating when the latter is dominated by the potential
density of a scalar field, called the inflaton field. Similarly, dark energy
can also be modelled as a suitable scalar field, called quintessence
\cite{quint}. It is natural to attempt to unify the two phases and consider that
accelerated expansion in the Universe is due to a single agent. The proposal is
called quintessential inflation \cite{QI}.

Apart from being economic, quintessential inflation addresses holistically
accelerated expansion in the early and late Universe in a single theoretical
framework. A successful quintessential inflation model has to satisfy the
observations of both inflation and dark energy. As such, constructing
a quintessential inflation model is highly constrained and very difficult to
achieve, but not impossible (e.g. see
Ref.~\cite{QIlatest1,QIlatest2,QIlatest3,QIlatest4,QIlatest5}
for recent reviews).

From the very beginning, modelling cosmic inflation was attempted in modified
gravity as well as particle physics. Indeed, the very first inflation model is
Starobinsky's $R^2$ inflation \cite{staro}. It is harder to use modified gravity
for dark energy however, because deviation from Einstein's general relativity
should not violate stringent constraints set by a plethora of experiments
(solar system, E\"{o}tv\"{o}s etc.). This is why, in attempting to construct
a quintessential inflation model, we assume a blended approach, where modified
gravity is mainly employed for inflation, while particle theory (which is
behind our scalar potential) accounts for dark energy.

In our model, we consider the Palatini formulation of
gravity~\cite{Palatini1919,Ferraris1982}. In the Palatini
formulation the connection and the metric are independent variables. In general
relativity the traditional metric formalism and the Palatini one are equivalent.
However, this is not so when matter is non-minimally coupled to gravity or
when the action is no longer linear in $R$, the curvature scalar.

Metric $R^2$ gravity introduces a new degree of freedom (dof), which can be
expressed as a scalar field (scalaron) in the Einstein frame \cite{staro}.
In contrast, Palatini $R^2$ gravity, has no extra propagating dof that can
play the role of the inflaton field. Therefore an additional scalar field must
be introduced.

In Palatini $R^2$ inflation, one can lower the tensor-to-scalar ratio in any
scalar field inflation model~\cite{Enckell:2018hmo, Antoniadis:2018ywb}.
Moreover, Palatini modified gravity evades the
stringent constraints on the propagation speed of primordial gravitational
waves. Finally, Palatini gravity does not suffer that much from solar system
and other related bounds on modified gravity, which means it is ideal for
modelling quintessential inflation \cite{samuel}.

In quintessential inflation, the thermal bath of the hot big bang is not
generated by the decay of the inflaton field, because the latter must survive
until the present to become quintessence. An alternative mechanism for reheating
the Universe must be employed. In this work, we do not consider a specific
mechanism for reheating the Universe. However, there is a plethora of such
mechanisms \cite{instant,curvreh1,curvreh2,riccireh1,riccireh2,riccireh3} (see
also Ref.~\cite{otherreh1,otherreh2}),
and we assume the operation of one of them.

We use natural units, where \mbox{$c=\hbar=1$} and \mbox{$8\pi G=m_P^{-2}$}
with {$m_P=2.43\times 10^{18}\,$GeV} being the reduced Plank mass.

\section{The model}

We consider the action in the Palatini formalism
\begin{equation}
  S = \int{\rm d}^4 x \sqrt{-g} \left[ \frac12 m_P^2F(\varphi,R) -
    \frac12 g^{\mu\nu}\partial_\mu\varphi\partial_\nu\varphi - V(\varphi)\right]
  +S_m[g_{\mu\nu},\psi]\,,
    \label{eq:S_Jordan}
\end{equation}
where $\psi$ collectively represents the matter fields other than the
inflaton. %
The function $F(\varphi,R)$ takes the following form%
\footnote{See Ref.~\cite{Nojiri1,Nojiri2} for recent reviews on $F(R)$ gravity
  and Ref.~\cite{Oikonomou:2021msx} for a recent study of $F(\varphi,R)$
  phenomenology.}
\begin{equation}
  F(\varphi,R)=\left(1+\xi\frac{\varphi^2}{m_P^2}\right)R+
  \frac{\alpha}{2m_P^2}R^2\,,
    \label{F}
\end{equation}
with $R$ being the  Ricci scalar, which is a function of the connection only
\begin{equation}
    R=g^{\mu\nu}R_{\mu\nu}(\Gamma)\,.
\end{equation}

Note that both terms in Eq.~(\ref{F}) are well motivated in the literature since
they can naturally arise when one considers quantum corrections
(e.g. see Ref.~\cite{Codello:2015mba}).
The above action is dynamically equivalent to 
\begin{eqnarray}
\mbox{\hspace{-3cm}}
  S & = & \int{\rm d}^4 x \sqrt{-g}\left[\frac12m_P^2
    \left(1+\xi\frac{\varphi^2}{m_P^2}
    +\frac{\alpha}{m_P^2}\chi\right)R
    -\frac14\alpha\chi^2-
    \frac12g^{\mu\nu}\partial_\mu\varphi\partial_\nu\varphi -
    V(\varphi)\right]\nonumber\\
  && +S_m[g_{\mu\nu},\psi]\,,
\end{eqnarray}
where $\chi$ is an auxiliary scalar field, which will be dispensed below.

To assist our intuition, we switch to the Einstein frame by a suitable 
conformal transformation
\begin{equation}
  g_{\mu\nu}\rightarrow \bar{g}_{\mu\nu}=
  \left(1+\xi\frac{\varphi^2}{m_P^2} 
  +\frac{\alpha}{m_P^2}\chi\right)g_{\mu\nu} \, .
\end{equation}
Now we eliminate the auxiliary field by obtaining its equation of motion
\begin{equation}
  \frac{\delta{S}}{\delta{\chi}}=0 \quad \Leftrightarrow \quad
  \chi=\frac{4m_P^2V+(m_P^2+\xi\varphi^2)(\bar{\partial}\varphi)^2}{
(m_P^2+\xi\varphi^2)m_P^2-\alpha(\bar{\partial}\varphi)^2}\,,
\end{equation}
where \mbox{$(\bar{\partial}\varphi)^2\equiv
  \bar g^{\mu\nu}\bar\partial_\mu\varphi\bar\partial_\nu\varphi$}.
Substituting $\chi$ back into the action yields
\begin{eqnarray}
  \mbox{\hspace{-1cm}}
    && S_E=\int{\rm d}^4 x \sqrt{-\bar{g}}
  \left[\frac12m_P^2\bar{R}-\frac12({\partial}\varphi)^2
    \frac{1+\frac{\xi\varphi^2}{m_P^2}}{
      \left(1+\frac{\xi\varphi^2}{m_P^2}\right)^2+
      \frac{4\alpha V(\varphi)}{m_P^4}}\right.\nonumber\\
  \mbox{\hspace{-2cm}}
    && \left.-\frac{V(\varphi)}{\left(1+\frac{\xi\varphi^2}{m_P^2}\right)^2+
      \frac{4\alpha V(\varphi)}{m_P^4}}
    +\frac14\frac{\alpha}{m_P^4}
    \frac{({\partial}\varphi)^4}{
      \left(1+\frac{\xi\varphi^2}{m_P^2}\right)^2+
      \frac{4\alpha V(\varphi)}{m_P^4}}\right]
  +S_m[\Omega^{-2}\bar{g}_{\mu\nu},\psi]\,.
\label{quartic}
\end{eqnarray}
Note that in the Palatini formalism the auxiliary field is not dynamical, which
allowed us to use its equation of motion to eliminate it. Thus the resulting
action only contains one scalar field, albeit with non-canonical kinetic terms
and a modified potential. This is in contrast to the metric version of the
theory where the auxiliary field has its own kinetic term and the resulting
action is two-field.

The canonical field $\phi$ is obtained via the redefinition
\begin{equation} \label{phivarphi}
  \frac{{\rm d} \phi}{{\rm d} \varphi}=
  \sqrt{\frac{1+\frac{\xi\varphi^2}{m_P^2}}{
      \left(1+\frac{\xi\varphi^2}{m_P^2}\right)^2
      +\frac{4\alpha V(\varphi)}{m_P^4}}}\,.
\end{equation}
One can use $\varphi=\varphi(\phi)$ to obtain the potential in the Einstein
frame
\begin{equation} 
  \bar{V}(\phi)=\frac{V(\phi)}{\left(1+\frac{\xi\varphi^2(\phi)}{m_P^2}\right)^2
    +\frac{4\alpha V(\phi)}{m_P^4}}\,.
  \label{EinsteinV}
\end{equation}
The above suggests that, for very large values of $V(\phi)$ the term in
brackets in the denominator becomes negligible and the overall potential
in the Einstein frame approximates a constant given by
\mbox{$\bar V\rightarrow m_P^4/4\alpha$}. This is the inflationary plateau,
attained regardless of the specific form of $V(\varphi)$ as long as the latter
becomes very large in some limit.

Regarding the quintessential tail, the second flat region of the scalar
potential, which is responsible for dark energy at present, we note that
near the present time, $R$ is tiny, so the $\alpha R^2$ term in
the Lagrangian is negligible. This is equivalent to setting $\alpha=0$.
In this case, Eq.~(\ref{phivarphi}) reduces to
\begin{equation} \label{phivarphi0}
  \frac{{\rm d}\phi}{{\rm d}\varphi}=
  \frac{1}{\sqrt{1+\frac{\xi\varphi^2}{m_P^2}}} 
\end{equation}
which results in
\begin{equation}
\varphi=\frac{m_P}{\sqrt{\xi}}\sinh\left(\frac{\sqrt{\xi}\,\phi}{m_P}\right)\,.
\label{phiDE}
\end{equation}

We consider that the runaway potential of the inflaton/quintessence scalar
field $\varphi$ is 
\begin{equation}
    V(\varphi)=M^4\,e^{-\kappa\,\varphi/m_P}\,,
\label{Vvarphi}
\end{equation}
where the dimensionless constant $\kappa$ is the strength of the exponential
and $M$ is an energy scale. An exponential potential is well motivated in
particle physics. Using the above and
Eqs.~(\ref{EinsteinV}) and (\ref{phivarphi0}) with \mbox{$\alpha=0$}, we find
that the scalar potential of the quintessential tail in the Einstein frame is
\begin{equation} 
\bar V(\phi)=M^4\frac{\exp
  \left[-\frac{\kappa}{\sqrt{\xi}}
    \sinh\left(\frac{\sqrt{\xi}\,\phi}{m_P}\right)\right]}{
    \cosh^4\left(\frac{\sqrt{\xi}\,\phi}{m_P}\right)}\,.
\label{EinsteinV0}
\end{equation}
Note that, in the limit \mbox{$\sqrt\xi\,\phi\ll m_P$} the above reduces to
\mbox{$\bar V=M^4\exp(-\kappa\,\phi/m_P)$}. In this limit, Eq.~(\ref{phivarphi})
suggests \mbox{$\phi\approx\varphi$}, i.e. $\varphi$ is approximately
canonical. Thus, in this limit we end up with the usual exponential
quintessential tail, which leads to accelerated expansion only if
\mbox{$\kappa<\sqrt 2$}. When $\kappa$ is larger, the exponential potential
leads to the scaling solution which cannot result in accelerated expansion.
However, as $\phi$ grows, the Einstein frame potential becomes steeper than an
exponential and accelerated expansion ceases, even if $\kappa$ is small enough.

If $\kappa$ is small enough to lead to accelerated expansion when $\phi$ is
small, then inflation would not be able to end even after the field exits the
inflationary plateau.
This is why we consider the effect of the $\alpha R^2$ term,
which is negligible at late times, but important at early times and high
energies. Thereby, we can facilitate a graceful exit from inflation and still
achieve accelerated expansion at present. However, we find that the value of
the non-minimal coupling $\xi$ is not the same for successful inflation and
quintessence. Therefore, we consider a mild running of $\xi$ as follows
\begin{equation}
\xi(\varphi)=\xi_*\left[1+\beta\ln\left(\frac{\varphi^2}{\mu^2}\right)\right]\,,
    \label{xirun}
\end{equation}
where $\mu$ is a suitable mass scale, and $\xi_*$ and $\beta$ are constants,
to be determined by the observations. The above is suggested by renormalization
considerations. The scalar field only slowly varies
(rolls) when the cosmological scales exit the horizon during inflation and also
when quintessence thaws while dominating the Universe at present. This means
that, in both cases, \mbox{$\xi\simeq\,$constant}. However, because $\varphi$
changes dramatically between inflation and quintessence, the non-minimal
coupling is not going to be the same in both cases.

\begin{figure}[h]
\begin{center}
\includegraphics[scale=0.6]{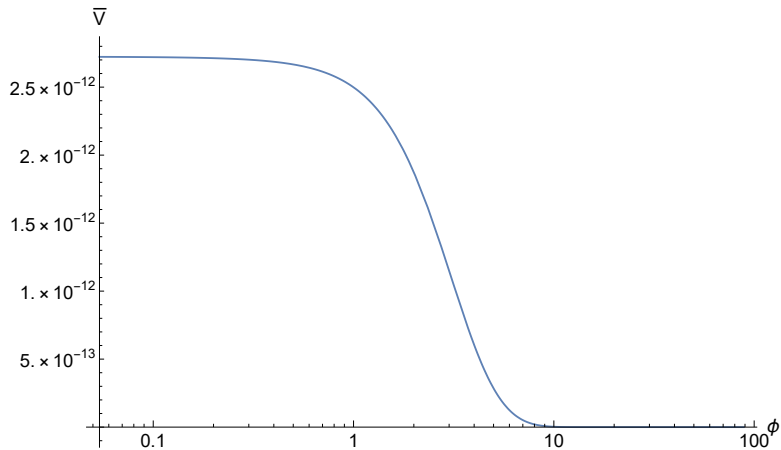}
\end{center}
\caption{The scalar potential $\bar V$ in Planck units in the Einstein frame,
  featuring the inflationary plateau and the quintessential tail.}
\label{EinV}
\end{figure}

The scalar potential in the Einstein frame is depicted in Fig.~\ref{EinV}.

\section{Equation of motion}

Varying the action with respect to $\varphi$, we have
\begin{equation} \label{KGvarphi}
  \ddot{\varphi}+3H\dot{\varphi}+V'(\varphi)
  -\left[\xi(\varphi)+\frac12 \xi'(\varphi)
    \varphi\right]\varphi R=0,
\end{equation}
which, using Eq.~(\ref{xirun}), immediately reads
\begin{equation}
\ddot{\varphi}+3H\dot{\varphi}+V'(\varphi)=\xi_*
\left[1+\beta\left(1+\ln{\frac{\varphi^2}{\mu^2}}\right)\right]\varphi R\,,
\end{equation}
where the prime denotes derivative with respect to the argument ($\varphi$ in
this case) and the dot
denotes derivative with respect to time in the Jordan frame.

To investigate $R$ we need to consider the energy-momentum tensor. We have
\begin{equation}
  T_{\mu\nu}=-\frac{2}{\sqrt{-g}}\frac{\delta S}{\delta g^{\mu\nu}}=
\left(F_{R}R_{(\mu\nu)}-\frac{1}{2}g_{\mu\nu}F\right)m_P^2\,,
\label{Tmn}
\end{equation}
where \mbox{$F_{R}(\varphi,R)\equiv \partial_R F(\varphi,R)$} with
$F(\varphi,R)$ given in Eq.~(\ref{F}). The trace of the
above is \mbox{$T=(F_{R}R-2F)m_P^2$}. Thus, the (Palatini) Ricci scalar is
algebraically related to the matter sources as
\begin{equation}
  R=\frac{1}{m_P^2+\xi\varphi^2}\left[\rho(1-3w)
  -\dot{\varphi}^2+4V(\varphi)\right]\,,
\label{RJF}
\end{equation}
where we used that the trace of the energy-momentum tensor is \cite{ours}
\begin{equation}
T=\dot{\varphi}^2-4V(\varphi)-\rho(1-3w)\,,
\label{Ttrace}
\end{equation}
with $\rho$ the energy density and $w$ the barotropic parameter of the
background matter,
dominant or not. Note that, when the background matter is dominant, then
\mbox{$T=(3w-1)\rho$}, which is zero during radiation domination, since then
$w=\frac13$. The same is true of $R$ itself. As a result, during radiation
domination the equation of motion of $\varphi$ reduces to the usual
Klein-Gordon of a minimally coupled scalar field.
It is also interesting that both $R$ and $T$ above are independent from the
value of $\alpha$.\footnote{This is because of the global scale invariance of
  the $R^2$ term, which is true in both the metric and the Palatini formalisms.
  We are thankful to the referee for pointing this out.}


In the Einstein frame, there is a new coupling between the matter action and
the inflaton field. Indeed, its equation of motion now reads \cite{ours}
\begin{equation}
  \frac{\delta S_E}{\delta\phi}=
  \sqrt{-\bar{g}}\left(\ddot\phi+3H\dot\phi\right)+
  \frac{{\rm d}\varphi}{{\rm d}\phi}
  \left(\sqrt{-\bar g}\bar V'(\varphi)+
  \frac{\delta S_m}{\delta\varphi}\right)=0\,,
  \label{dSE}
\end{equation}
where ${\rm d}\varphi/{\rm d}\phi$ is given by Eq.~(\ref{phivarphi}).
The functional derivative of the matter action is \cite{ours}
\begin{eqnarray}
  \frac{\delta S_m}{\delta\varphi}=
  \frac{\sqrt{-\bar g}\,\xi\varphi}
       {m_P^2+\xi\varphi^2-\alpha(\bar{\partial}\varphi)^2/m_P^2}
  \,\bar\rho(1-3\bar w)\,,
  \label{dSm}
\end{eqnarray}
where $\bar w$ is the barotropic parameter of the background matter
(assumed to be a barotropic ideal fluid), which is the same in both the Einstein
and the Jordan frames \mbox{$\bar w=w$} \cite{ours}. From the
above we see that, when the background matter is radiation (dominant or not),
for which \mbox{$w=\frac13$}, then the coupling of the inflaton to matter
disappears. Thus, this coupling is only effective after matter-radiation
equality, when the Universe is matter dominated. As we have discussed, on late
times the contribution of the $\alpha R^2$ term in the Lagrangian density is
negligible. This is equivalent to setting \mbox{$\alpha\rightarrow 0$} in the
above.

Regarding the derivative of the potential in Eq.~(\ref{dSE}) (but neglecting
the running of $\xi$ as subleading), we have
\begin{equation}
  \frac{{\rm d}\bar{V}(\varphi)}{{\rm d}\varphi}=
  \frac{V'(\varphi)}{\left(1+\frac{\xi\varphi^2}{m_P^2}\right)^2+
    \frac{4\alpha V}{m_P^4}}-
  \frac{V(\varphi)\left[\frac{4\alpha}{m_P^4}V'(\varphi)+
      \frac{4\xi\varphi}{m_P^2}\left(1+\frac{\xi\varphi^2}{m_P^2}\right)\right]}
       {\left[\left(1+\frac{\xi\varphi^2}{m_P^2}\right)^2+
           \frac{4\alpha V}{m_P2^4}\right]^2}\,,
\end{equation}
where we considered Eq.~(\ref{EinsteinV}).

\section{Inflation}

Inflation is expected to occur when we are on the inflationary plateau
(in the Einstein frame) with a large value of $V$, i.e. \mbox{$\varphi\ll 0$}.
In this limit, we can consider slow-roll inflation in the Einstein frame, which
is determined by the slow-roll parameters
\begin{equation} \label{SR}
  \epsilon\equiv\frac12
  \left(\frac{{\rm d}\bar{V}}{{\rm d}\phi}\frac{m_P}{\bar{V}}\right)^2
\quad{\rm and}\quad \eta\equiv 
   \frac{{\rm d}^2\bar{V}}{{\rm d}\phi^2}\frac{m_P^2}{\bar{V}} \,.
\end{equation}
To have slow-roll, \mbox{$\epsilon<1$} and \mbox{$|\eta|<1$}.
Ref.~\cite{Enckell:2018hmo} suggests that the above are given by
\begin{eqnarray}
\mbox{\hspace{-2cm}}
  \epsilon = \frac{\tilde\epsilon}{1+4\alpha\tilde V} & {\rm with} &
  \tilde\epsilon=\frac12
  \frac{\left[\kappa\left(1+\frac{\xi\varphi^2}{m_P^2}\right)
      +4\xi\frac{\varphi}{m_P}\right]^2}{1+\frac{\xi\varphi^2}{m_P^2}}\,,
  \label{epsV}\\
\mbox{\hspace{-1.5cm}}
  \eta = \tilde\eta - 3\frac{4\alpha\tilde V}{1+4\alpha\tilde V} &
      {\rm with} & 
      \tilde\eta=\frac{\left(7\kappa\xi\frac{\varphi}{m_P}
        +\kappa^2\right)\left(1+\frac{\xi\varphi^2}{m_P^2}\right)
        -4\xi+16\xi^2\frac{\varphi^2}{m_P^2}}{1+\frac{\xi\varphi^2}{m_P^2}}\,,
\label{etaV}
\end{eqnarray}
where the tilded quantities correspond to \mbox{$\alpha=0$}
(and we have taken the limit of constant $\xi$).
In the above
\begin{equation}
  \tilde V\equiv
  \frac{M^4\,e^{-\kappa\varphi/m_P}}{(1+\frac{\xi\varphi^2}{m_P^2})^2}\,
  \label{tildeV}
\end{equation}
which is readily obtained by Eqs.~(\ref{EinsteinV}) and 
(\ref{Vvarphi}) when \mbox{$\alpha\rightarrow 0$}. To contrast with observations
we obtain the standard inflationary observables
\begin{eqnarray}
&&    
  n_s=1-6\epsilon+2\eta=1-6\tilde\epsilon+2\tilde\eta
  \, , \label{ns}\\
&&
  r=16\epsilon
  \quad{\rm and}\quad 
24\pi^2 m_P^4 A_s=\frac{\bar V}{\epsilon}=\frac{\tilde{V}}{\tilde\epsilon}\,,
\label{rAs}
\end{eqnarray}
where $A_s$ is the scalar power spectrum, $n_s$ is the spectral index and $r$ is
the tensor-to-scalar ratio at the CMB pivot scale $k_*=0.05\,{\rm Mpc}^{-1}$.
In the above, we used that \mbox{$\bar V/\epsilon=\tilde V/\tilde\epsilon$}
as shown in Ref.~\cite{Enckell:2018hmo}.
Technically, \mbox{$\beta\neq 0$}, so Eqs.~(\ref{ns}) and (\ref{rAs})
are not exact, but we expect the modification to be minor because
the non-minimal coupling depends only logarithmically on the slowly rolling
inflaton (cf. Eq.~(\ref{xirun})).

The observations suggest \cite{Planck,BICEP}
\begin{equation} \label{CMBobs}
  \ln\left(10^{10} A_s\right)=3.044\pm0.014\, , \quad
  n_s=0.9649\pm0.0042\, , \quad 
  r < 0.036 \, .
\end{equation}
From this and Eq.~(\ref{rAs}), it is straightforward to find 
\begin{equation}
  2.1\times 10^{-9}=A_s=
  \frac{M^4\,e^{-\kappa\varphi_*/m_P}}{12\pi^2m_P^4
    \left(1+\frac{\xi\varphi^2_{*}}{m_P^2}\right)\left[\kappa\left(
      1+\frac{\xi\varphi^2_{*}}{m_P^2}\right)+4\xi\varphi_{*}\right]^2}\,,
    \label{adfbadf}
\end{equation}
where the subscript `$*$' denotes the exit of the pivot scale during inflation
and we employed Eqs.~(\ref{epsV}) and (\ref{tildeV}).
From Eqs.~(\ref{epsV}), (\ref{etaV}) and (\ref{ns}), the spectral index is
\begin{equation}
  n_s-1=-\kappa^2\left(1+\frac{\xi\varphi_*^2}{m_P^2}\right)
  -10\,\xi\kappa\frac{\varphi_*}{m_P}
  -8\xi\;\frac{1+2\left(\frac{\xi\varphi_*^2}{m_P^2}\right)}
  {1+\frac{\xi\varphi_*^2}{m_P^2}}\,.
\end{equation}
Finally, for the number of e-folds we have
\begin{eqnarray}
  & N & =-\frac{1}{m_P}\int\frac{{\rm d}\phi}{\sqrt{2\epsilon}}=
  -\frac{1}{m_P}\int\frac{{\rm d}\varphi}
  {\kappa\left(1+\frac{\xi\varphi^2}{m_P^2}\right)
    +4\,\frac{\xi\varphi}{m_P}}\nonumber\\
  &&=-\frac{1}{\sqrt{\xi\kappa^2-4\xi^2}}\arctan
  \left[\frac{\sqrt\xi\left(2+\kappa\frac{\varphi}{m_P}\right)}
    {\sqrt{\kappa^2-4\xi}}\right].
\end{eqnarray}

\section{Kination}

\begin{figure}[h]
\begin{center}
\includegraphics[scale=0.5]{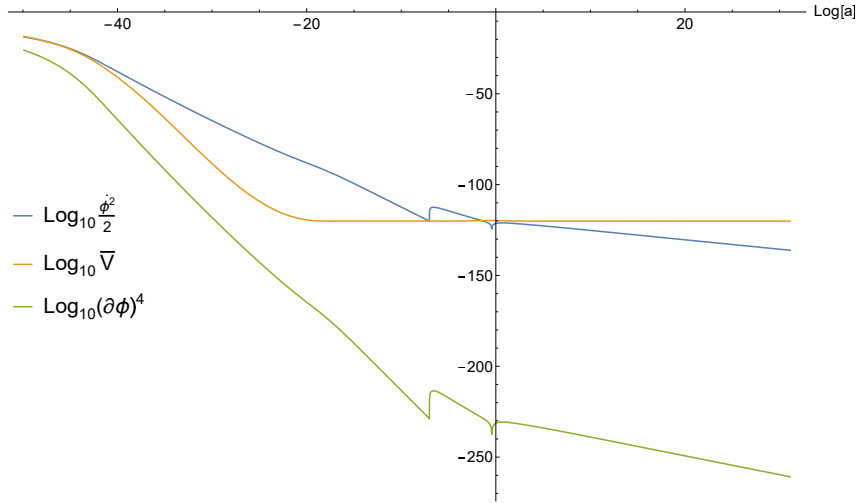}
\end{center}
\caption{Log-log plot in Planck units
  of the contributions to the inflaton energy density as a
  function of the scale factor normalised to unity at present, starting at the
  end of inflation. The upper curve (blue) corresponds to the
  canonical kinetic energy density of the scalar field, which dominates the
  potential until the time of equal-matter-radiation densities (equality). At
  this time the field briefly freezes only to unfreeze in matter domination
  (because of the interaction with matter in Eq.~(\ref{dSm})) and become again
  dominated by its kinetic energy density until the present time when the
  potential energy density, depicted by the middle curve (orange), takes over.
  The quartic kinetic term in Eq.~(\ref{quartic}), depicted by the lower curve
  (green), remains always negligible.}
\label{energies}
\end{figure}

At some point the inflationary plateau ends and the potential becomes steep and
curved. Inflation ends and the inflaton field falls over a potential cliff. As
a result, the kinetic energy density of the field dominates the Universe. The
Palatini Ricci tensor in Eq.~(\ref{RJF}) becomes
\mbox{$R=-\dot\varphi^2/(m_P^2+\xi\varphi^2)$}. In principle, the quartic
kinetic term in Eq.~(\ref{quartic}) might affect the dynamics of kination, but
we find otherwise (see Fig.~\ref{energies}).
Also, as we have seen, the coupling of the inflaton to the
background matter disappears if the background is radiation.
Thus, kination proceeds as usual, with
\mbox{$\rho\propto a^{-6}$} and \mbox{$a\propto t^{1/3}$}
\cite{kination}.


We assume that subdominant radiation is generated at the end of inflation
(denoted by the super/subscript `end'),
with density parameter
\mbox{$\Omega_r^{\rm end}=(\rho_r/\rho_{\rm tot})_{\rm end}$}, which
is also called the reheating efficiency.
The density of the background radiation scales as \mbox{$\rho_r\propto a^{-4}$}.
This means that there is a moment when $\rho_r$ becomes dominant over the
rolling scalar field and the Universe becomes radiation dominated. This is the
moment of reheating. After reheating, the field continues to roll kinetically
dominated for a while until its potential density becomes important. If the
slope of the latter is small enough the field freezes.

\begin{figure}[h]
\begin{center}
\includegraphics[scale=0.5]{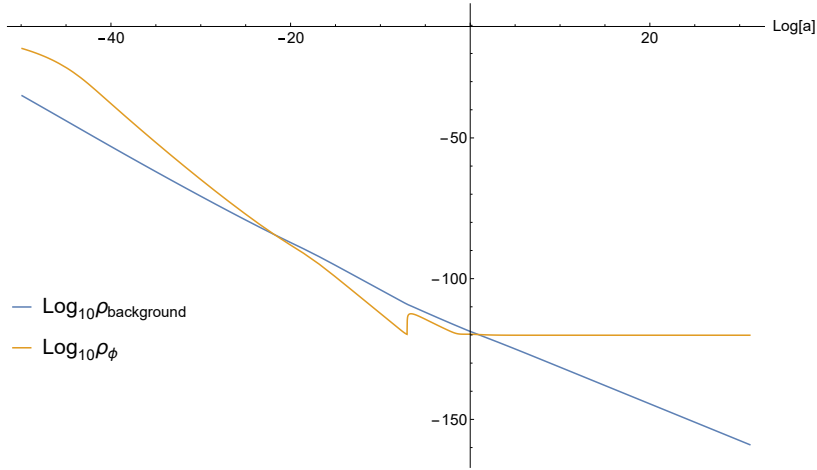}
\end{center}
\caption{Log-log plot of the energy density of the scalar field in Planck units
  (orange) and that of the background (blue) after the end of inflation.
  Originally, the scalar field kinetic density dominates (kination) until the
  moment of reheating when the background density takes over. The scalar field
  momentarily freezes at equality, but then unfreezes in the matter era, due to
  the background backreaction. Eventually, it comes to dominate at present.}
\label{energybackground}
\end{figure}

Things change after matter-radiation equality, when the interaction of the
field (which is now quintessence) with matter affects its dynamics. We find
that quintessence unfreezes and rolls further, until it dominates the Universe
again. The evolution of the energy density of the scalar field and of the
background density is shown in Fig.~\ref{energybackground}.

The early, stiff kination era, increases the number of e-folds between the end
of inflation and the horizon exit of the CMB scale from the standard 50–60 to
60–75. We have taken the full expansion history into account when fixing
the CMB scale in our results.

\section{Quintessence}

Soon after matter-radiation equality, quintessence refreezes at some value
$\phi_0$ (or $\varphi_0$ in terms of the non-canonical field). Then there are
certain requirements it must satisfy if it is to be the observed dark energy,
akin to the CMB observational constraints for inflation. The first such
constraint is Coincidence. This means that the density parameter of the frozen
quintessence at present must be \cite{planck2}
 \begin{equation}
   \Omega_{\phi}=\Omega_{\mbox{\tiny DE}}=0.6847\pm 0.0073\,.
\label{coincidence}
 \end{equation}
 In general, the barotropic parameter of quintessence is variable. By Taylor
 expanding it near the present, this varying barotropic parameter can be 
approximated as (CPL parametrisation, \cite{CPL1,CPL2})
 \begin{equation}
     w_{\mbox{\tiny DE}}=w_{\mbox{\tiny DE}}^{0}+w_{\rm a}\left(1-\frac{a}{a_0}\right),
 \end{equation}
 where $w_{\mbox{\tiny DE}}^{0}$ is the value of $w_{\mbox{\tiny DE}}$ at present and
 \begin{equation}
   w_{\rm a}=-\left.\frac{{\rm d}w_{\mbox{\tiny DE}}}{{\rm d}a}\right|_{a_0}=
-\left.\frac{{\rm d}w_{\mbox{\tiny DE}}}{{\rm d}t}\frac{1}{\dot{a}}\right|_{t_0}\;,
   \label{CPL}
 \end{equation}
 where `0' denotes the present time. Observations require \cite{planck2}
\begin{equation}
  -1\leq w_{\mbox{\tiny DE}}^{0}<-0.95
  \quad{\rm and}\quad
 w_{\rm a}\in[-0.55,0.03]\,.
\label{w0obs}
\end{equation}
Demanding that quintessence is successful dark energy implies that
\mbox{$w_\phi=w_{\mbox{\tiny DE}}$}, which must satisfy the above constraints.

Starting with the coincidence requirement, the quintessence density at present
is
\begin{equation}
  \rho_\phi^0 = 3H_0^2 m_P^2 \Omega_\phi=3H_0^2 m_P^2 \Omega_{\mbox{\tiny DE}}
  \approx 8\times10^{-121} m_P^4 \, ,
\end{equation}
where we approximated \mbox{$H_0 \approx 67.8$}~km/s/Mpc and we used
Eq.~(\ref{coincidence}). Eq.~(\ref{tildeV}) suggests
\begin{equation} \label{eq:rho_DE}
  \rho_\phi^0\simeq\bar V(\varphi_0)=\frac{M^4\,e^{-\kappa \varphi_0/m_P}}
  {\left(1+\frac{\xi\varphi_0^2}{m_P^2}\right)^2}
  =M^4\,\frac{e^{-\frac{\kappa}{\sqrt\xi}\sinh(\sqrt\xi \phi_0/m_P)}}
  {\cosh^4(\sqrt\xi \phi_0/m_P)}\,,
  \label{rhophi0}
\end{equation}
where we considered Eq.~(\ref{EinsteinV0}) because $\alpha$ is negligible at
late times. In the above $\xi$ is not the same as in inflation, but it is given
by Eq.~(\ref{xirun}) as \mbox{$\xi=\xi_*[1+\beta\ln(\varphi_0^2/\mu^2)]$}.
We have also taken into account that the field is thawing so that its kinetic
energy density is subdominant to its
potential energy density and so \mbox{$\rho_\phi\simeq V$}. Because $\xi$ is
logarithmicaly dependent on $\varphi$ and the later varies mildly as
quintessence thaws, we expect \mbox{$\xi\simeq\,$constant}.

The value of $M$ is  determined by the normalisation of the scalar
perturbations during inflation:
\begin{equation} \label{eq:As}
    A_s = \frac{2V(\varphi_*)}{3\pi^2m_P^4 r} \, .
\label{V0norm}
\end{equation}

We further consider \mbox{$|\xi|=|\xi(\varphi_0)|\ll 1$}. In this limit,
Eqs.~(\ref{rhophi0}) and (\ref{V0norm}) suggest
\begin{equation} 
  \frac{\kappa\varphi_0}{m_P}=
  -\left[\ln\left(\frac{\rho_\phi^0}{m_P^4}\right)
    -\ln\left(\mbox{$\frac{3\pi^2}{2}$}A_s r\right)\right]
  +\frac{\kappa\varphi_*}{m_P}
  \approx 252+\ln\left(\frac{r}{10^{-3}}\right)+ \frac{\kappa\varphi_*}{m_P}
  \, .
    \label{varphi0}
\end{equation}
Because we find that \mbox{$\kappa\varphi_*\sim-(\mbox{a few})\times m_P$} we
expect \mbox{$\kappa\varphi_0/m_P\simeq\,$250} or so.
Using Eq.~(\ref{phiDE}), we
find
\begin{equation}
\frac{\sqrt\xi\,\phi_0}{m_P}\simeq
\sinh^{-1}\left\{\frac{\sqrt\xi}{\kappa}
  \left[252+\ln\left(\frac{r}{10^{-3}}\right)\right]
  +\sinh\left(\frac{\sqrt\xi\,\phi_*}{m_P}\right)\right\}\,.
\label{phi0}
\end{equation}

\section{Results}

\begin{figure}[h]
\begin{center}
  \includegraphics[scale=0.3]{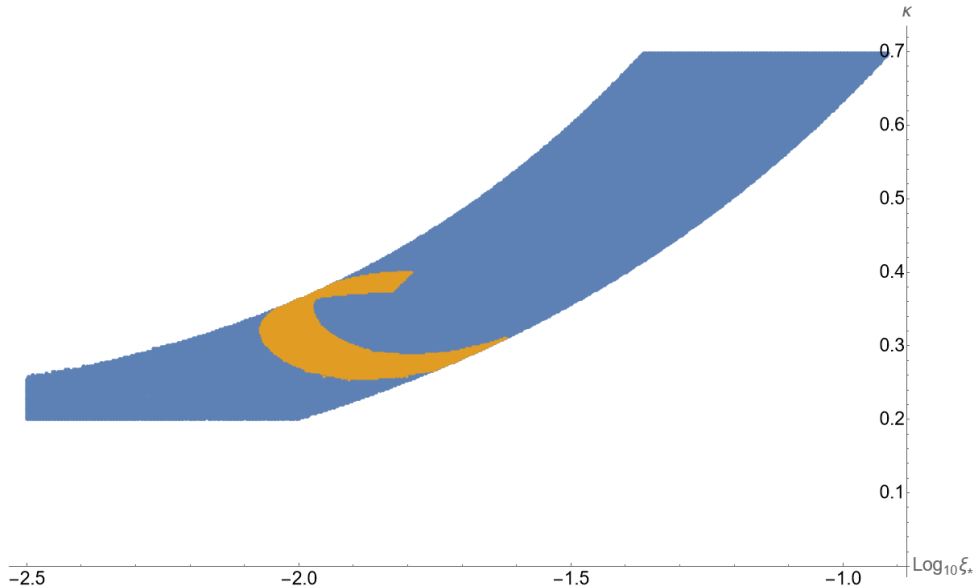}
\end{center}
\caption{The parameter space $\kappa(\xi_*)$ for successful inflation
  The blue (dark) band depicts the region which reproduces the observed
  values of the spectral index and the amplitude of the cosmological
  perturbations (the central values of $n_s$ and $A_s$ in Eq.~(\ref{CMBobs})).
  The allowed region is depicted in orange (light band), which satisfies the
  bound on the tensor-to-scalar ratio $r$ in Eq.~(\ref{CMBobs}) and
  corresponds to the range of reheating efficiency in Eq.~(\ref{Orange}),
  which in turn implies the number of efolds of remaining inflation
  when the cosmological scales leave the horizon ranges as
  \mbox{$N_*=\,$60--75}. (We have taken \mbox{$\alpha M^4=143.08$})}
\label{param}
\end{figure}

\begin{table}
\begin{center}  
\begin{tabular}{lll}\hline
  $\alpha=9.16\times 10^{10}$ &
  $\kappa=0.2956$ &
  $M^4=2.11\times 10^{-9}\,m_P^4$\\
  $\xi_*=0.0093282$ & $\beta=-0.10075$ & $\mu=-6\,m_P$\\\hline
\end{tabular}
\end{center}
\caption{Exact values of model parameters assumed in all the figures.}
\label{table}
\end{table}  

The parameter space for successful inflation is shown
in Fig.~\ref{param}. From this figure it is evident that inflation requires
(see Table.~\ref{table}, for exact values used in the figures).
\begin{equation}
\kappa\approx 0.3
  \quad{\rm and}\quad
\xi_*\approx 0.01
\label{kappaxi}
\end{equation}  
where, without loss of generality, we chose that
\mbox{$\mu^2\approx\varphi_*^2$} in Eq.~(\ref{xirun})
such that, when the cosmological scales leave
the horizon, we have \mbox{$\xi\approx\xi_*$}.
We find that \mbox{$\mu\simeq -6\,m_P$}
(for \mbox{$\mu=-6.00\,m_P$} we find \mbox{$\varphi_*=-5.91\,m_P$}).
For the $\alpha$ parameter, we obtain a lower
bound \mbox{$\alpha\gtrsim 10^7$}. We choose \mbox{$\alpha\simeq 10^{11}$}.
The energy scale at the end of inflation is found to be
\mbox{$\bar V_{\rm end}^{1/4}\simeq 3\times 10^{-5}\,m_P\sim 10^{14}\,$GeV}, which is
somewhat smaller than the estimate of the inflationary plateau
\mbox{$\bar V^{1/4}=(4\alpha)^{-1/4}m_P\sim 10^{-3}\,m_P$}. Similarly, the density
scale in the scalar potential is \mbox{$M^4\simeq 2\times 10^{-9}m_P^4$} which
implies \mbox{$M=2\times 10^{16}\,$GeV}, i.e. the scale of grand unification.

For successful quintessence we consider the reheating efficiency
\mbox{$\Omega_r^{\rm end}
  \sim 10^{-15}$}. This value belongs comfortably in the allowed range,
\begin{equation}
  10^{-2}\left(\frac{\bar H_{\rm end}}{m_P}\right)^2\sim
  10^{-20}<\Omega_r^{\rm end}<1\,,
\label{Orange}
\end{equation}
where the upper bound corresponds to prompt reheating, while the lower bound
corresponds to gravitational reheating, for which
\mbox{$\rho_r^{\rm end}\sim 10^{-2}\bar H_{\rm end}^4$} \cite{ford1,ford2}.
In Eq.~(\ref{Orange}) we used
\mbox{$\bar H_{\rm end}\simeq\sqrt{(\bar V_{\rm end}/3)}/m_P\sim 10^{-9}\,m_P$}.
There are many possibilities for reheating the Universe without the decay of the
inflaton field, which are typically considered in non-oscillatory inflationary
models. Examples are instant preheating~\cite{instant}, curvaton
reheating~\cite{curvreh1,curvreh2} and Ricci
reheating~\cite{riccireh1,riccireh2,riccireh3}.

Let us estimate the reheating temperature. Assuming proper kination begins
right away after the end of inflation we find the following.
During kination, the total energy density of the Universe decreases as
\mbox{$\rho_{\rm tot}\simeq\rho_\phi\propto a^{-6}$}, while for radiation we have
\mbox{$\rho_r\propto a^{-4}$}, which means that
\mbox{$\rho_r/\rho_{\rm tot}\propto a^2$}.
Therefore,
\begin{equation}
  \Omega_r^{\rm end}=
  \left.\frac{\rho_r}{\rho_{\rm tot}}\right|_{\rm end}=
  \left(\frac{a_{\rm end}}{a_{\rm reh}}\right)^2
  \left.\frac{\rho_r}{\rho_{\rm tot}}\right|_{\rm reh}\simeq
  \left(\frac{a_{\rm end}}{a_{\rm reh}}\right)^2 \, ,
\label{aratio}
\end{equation}
where `reh' denotes reheating, which is the moment that radiation takes over
and we have \mbox{$\rho_r\simeq\rho_{\rm tot}$}. The density of the Universe at
reheating is straightforward to find, by considering that
\mbox{$\rho_{\rm tot}\propto a^{-6}$}. Indeed, we get
\begin{equation}
  \rho_{\rm reh}=\left(\frac{a_{\rm end}}{a_{\rm reh}}\right)^6\rho_{\rm end}
  \simeq 
  (\Omega_r^{\rm end})^3\,\bar V_{\rm end}\;,
\label{rhoreh}
\end{equation}
where we used Eq.~(\ref{aratio}) and that
\mbox{$\rho_{\rm end}\simeq\bar V_{\rm end}$}. Therefore, using that at reheating
\mbox{$\rho\simeq\rho_r=\frac{\pi^2}{30}g_* T^4$}, the reheating temperature is 
\begin{equation}
  T_{\rm reh}\simeq
  \mbox{$\frac{1}{\sqrt\pi}\left(\frac{30}{g_*}\right)^{1/4}$}
(\Omega_r^{\rm end})^{3/4}\,\bar V_{\rm end}^{1/4}\;,
\label{Treh}
\end{equation}
where $g_*$ is the number of effective relativistic degrees of freedom at
reheating. Putting in the numbers, we find \mbox{$T_{\rm reh}\sim 1\,$TeV}.
However, Figs.~\ref{energies} and \ref{energybackground} suggest that,
immediately after inflation, the energy density of the field does not fall as
rapidly as $a^{-6}$. This means that the radiation density takes over after the
above estimate, corresponding to a somewhat smaller reheating temperature.

\begin{figure}[h]
\begin{center}
\includegraphics[scale=0.6]{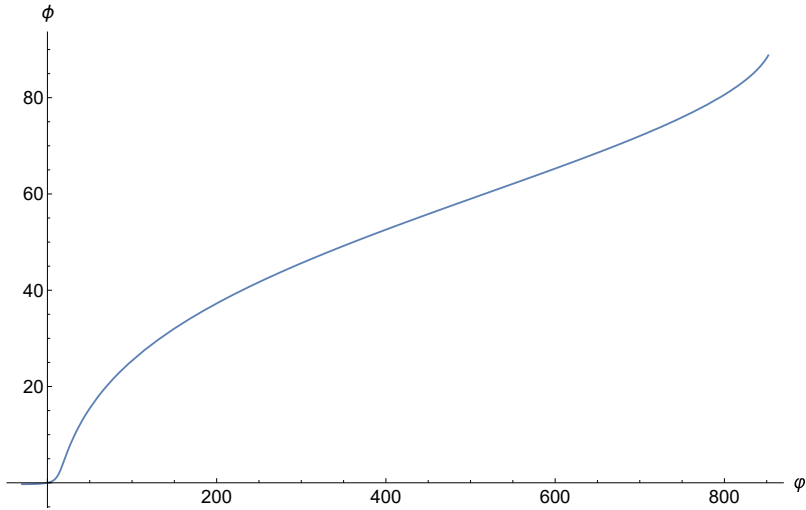}
\end{center}
\caption{The relation of the canonical $\phi$ with the non-minimal $\varphi$
when \mbox{$\kappa\simeq 0.3$} and \mbox{$\xi\sim 10^{-5}$}.}
\label{phivarphifig}
\end{figure}

The appropriate $\xi$ so that we can have successful coincidence is
\mbox{$\xi\sim 10^{-5}$}. In order for the running in Eq.~(\ref{xirun}) to
result in this value we find that we need
\mbox{$\beta\approx -0.101$}, which is rather reasonable.
From Eq.~(\ref{varphi0}) we can estimate
\mbox{$\varphi_0/m_P\approx e^{-1/2\beta}|\mu|\simeq 840$}. Then,
Eq.~(\ref{phiDE}) suggests \mbox{$\phi_0/m_P\simeq 80$}, as can be seen also in
Fig.~\ref{phivarphifig}.

With these values we see that \mbox{$\sqrt\xi\,\phi_0/m_P\simeq 0.25<1$}.
According to the discussion after Eq.~(\ref{EinsteinV0}), the potential
approximates a decaying exponential of strength $\kappa$. Since
\mbox{$\kappa<\sqrt 2$}, quintessence will approach the dominant attractor
solution, for which the barotropic parameter is \mbox{$w_\phi=-1+\kappa^2/3$}
\cite{DEbook}. With $\kappa=0.3$ we get \mbox{$w_\phi=-0.97$}.

\begin{figure}[h]
\begin{center}
\includegraphics[scale=0.6]{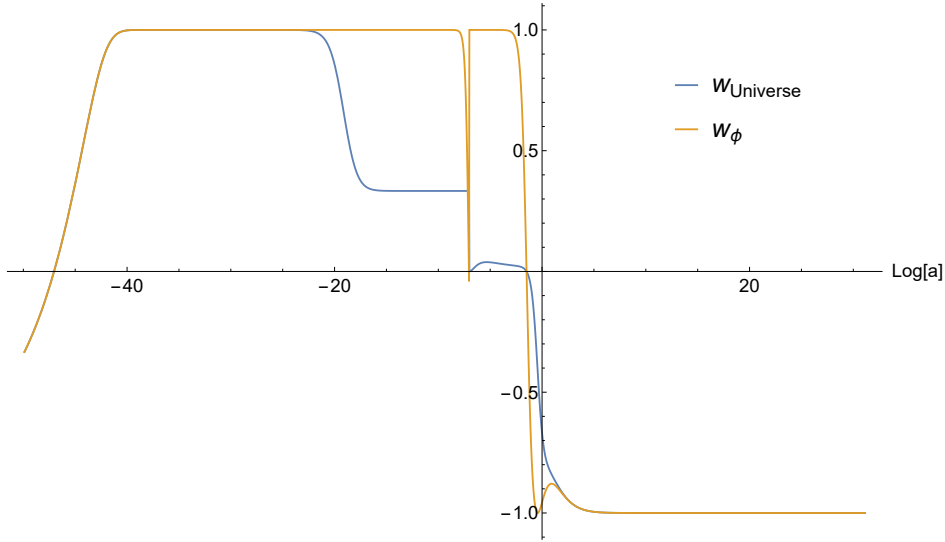}
\end{center}
\caption{The evolution of the barotropic parameters of the scalar field
  (orange) and of the Universe (blue) after the end of inflation. We see that
  during kination both barotropic parameters are the same and equal to unity,
  because the scalar field dominates. After reheating, the barotropic
  parameter of the Universe reduces to 1/3, while the scalar field continues to
  be kinetically dominated with barotropic parameter \mbox{$w_\phi=1$}. The
  field freezes briefly at radiation--matter equality, when its barotropic
  parameter is drastically reduced, while the Universe's barotropic parameter
  approximates zero. The backreaction due to matter does not allow the scalar
  field to stay frozen. Instead, it free-falls again with \mbox{$w_\phi=1$}
  until the present time, when it reduces drastically towards -1. The barotropic
  parameter of the Universe at present is found to be -0.669.}
\label{barotropics}
\end{figure}

However, the approximation is not very good because $\sqrt\xi\,\phi_0/m_P$ is
not very small. Indeed, using the parameter values in Table~\ref{table},
for the dark energy barotropic parameter today we find
\begin{equation}
w_\phi^0=w_{\mbox{\tiny DE}}^0=-0.956\quad{\rm and}\quad
w_{\rm a}=-0.1596\,,
\label{wresults}
\end{equation}
which satisfy the requirements in Eq.~(\ref{w0obs}) and will be observable in
the near future. The above is an existence proof that our model works. We
will attempt an exploration of the parameter space (which is a subset of the
one shown in Fig.~\ref{param}) in Ref.~\cite{ours}.
The evolution of the barotropic parameters of the scalar field
and the Universe after inflation is shown in Fig.~\ref{barotropics}.

\begin{figure}[h]
\begin{center}
\includegraphics[scale=1.1]{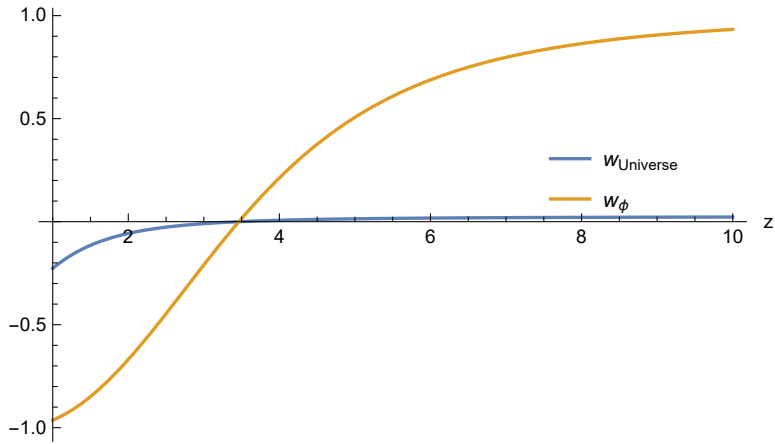}
\end{center}
\caption{The evolution of the barotropic parameters of the scalar field
  (orange) and of the Universe (blue) as a function of redshift near the
  present. It is evident that \mbox{$w\approx 0$} near \mbox{$z\simeq 4$},
  when galaxy formation occurs.}
\label{zoomz}
\end{figure}

We see that the barotropic parameter of the Universe after equality is not
exactly zero. In fact, we find that it peaks to almost \mbox{$w\simeq 0.04$} at
\mbox{$z\simeq 200$}. However, it reduces substantially for smaller redshifts
and is very close to zero near \mbox{$z\simeq\,$3--5}, which is when galaxy
formation occurs, as shown in Fig.~\ref{zoomz}.
It would be interesting to investigate characteristic observational signatures
of our scenario with respect to the growth of structures, but this is beyond
the scope of this paper.

\section{Conclusions}

We have investigated a model of quintessential inflation in the context of
Palatini modified gravity. We considered a non-minimally coupled scalar field
and an $R^2$ contribution to the Lagrangian, both of which are rather modest
modifications of gravity, frequently considered in the literature. The scalar
potential of our non-minimal field is simply an exponential, which is well
motivated in fundamental theory. The non-minimal coupling follows a mild
logarithmic running, expected by renormalisation considerations, such that it
is not the same during inflation and the present.

We find that our model can indeed successfully account for the observations of
inflation and dark energy without any unphysical fine-tuning. The strength of
the exponential potential is \mbox{$\kappa\simeq 0.3$} and the non-minimal
coupling runs from \mbox{$\xi\sim 10^{-2}$} during inflation to
\mbox{$\xi\sim 10^{-5}$} during quintessence. The non-perturbative coupling
of quadratic gravity is \mbox{$\alpha\gtrsim 10^7$} (we consider
\mbox{$\alpha\sim 10^{11}$}). The energy scale in our exponential potential
turns out to be \mbox{$M\simeq 2\times 10^{16}\,$GeV}, i.e. the scale of grand
unification. The barotropic parameter of quintessence and its
running are to be probed in the near future, e.g. by the EUCLID \cite{euclid}
and Nancy Grace Roman (former WFIRST) \cite{NGR1,NGR2} satellites.

Our model leads to a long period of kination (with reheating temperature 
\mbox{$T_{\rm reh}\lesssim 1\,$TeV}). After kination the field freezes but soon it
unfreezes again after equality (between matter and radiation), when the
back-reaction from a coupling to matter kicks in. We find that the barotropic
parameter of the matter era is affected in a diminishing way, such that it is
approximately zero at the time of galaxy formation, as required. It is an open
question whether its early values (almost 4\% at redshift 200 or so) affect
structure formation, in ways which could be an observational signature for our
scenario.

\section*{Acknowledgements}

KD is supported (in part) by the Lancaster-Manchester-Sheffield Consortium
for Fundamental Physics under STFC grant: ST/T001038/1.
AK is supported by the Estonian Research Council grant PSG761,
and ET by PRG803, MOBTT5, and PRG1055. AK and ET are both supported by the EU
through the European Regional Development Fund
CoE program TK133 ``The Dark Side of the Universe."
SSL is supported by the FST of Lancaster University.

\end{document}